# PUZZLING MAGNETO-OPTICAL PROPERTIES OF ZnMnO FILMS

M. Godlewski[1,2], A. Wąsiakowski[2], V.Yu. Ivanov[1], A. Wójcik-Głodowska[1], M. Łukasiewicz[2], E. Guziewicz[1], R. Jakieła[1], K. Kopalko[1], A. Zakrzewski[1]

[1] Institute of Physics, Polish Acad. of Sciences, Al. Lotników 32/46, 02-668 Warsaw, Poland

[2] Dept. Mathematics and Natural Sciences College of Science, Cardinal S. Wyszyński University, Warsaw, Poland

Y. Dumont

Groupe d'Etudes de la Matiere Condensee, CNRS- Universite de Versailles, 45 avenue des Etats-Unis, 78035 Versailles, France

**Abstract**

Optical and magneto-optical properties of ZnMnO films grown at low temperature by Atomic Layer Deposition are discussed. A strong polarization of excitonic photoluminescence is reported, surprisingly observed without splitting or spectral shift of excitonic transitions. Present results suggest possibility of Mn recharging in ZnO lattice. Strong absorption, with onset at about 2.1 eV, is related to Mn 2+ to 3+ photo-ionization. We propose that the observed strong circular polarization of excitonic emission is of a similar character as the one observed by us for ZnSe:Cr.

**Keywords**: Spintronics, low temperature growth, zinc oxide, optical properties

## 1. Introduction

Thin films of ZnTMO (TM = Mn, Co, Cu, Ni, Cr) are presently intensively studied for spintronics applications as possible room temperature ferromagnetic (RT FM) materials [1]. RT FM was predicted for ZnMnO films with a few percent of Mn concentration and with a very high free holes concentration of $10^{20}$ cm$^{-3}$ [2]. The latter condition is difficult to be achieved since as-grown ZnO is commonly of n-type [1]. Moreover, as discussed in this work, it is very unlikely to achieve a high p-type conductivity of ZnO with Mn staying in 2+ charge state.

Surprisingly, RT FM of ZnMnO was reported even for n-type samples (see [3-7] and references given there). It is now believed that the RT FM reported in these papers was due to inclusions of various TM oxides [4,8] and to metal accumulations and not due to carriers (holes) induced FM. Very importantly, such inclusions and accumulations, present in most of ZnMnO films grown with high temperature methods, can be avoided in films grown at low temperature (LT) [5-7,9].

In the previous study we have found that thin films of ZnMnO, when grown at low temperature by Atomic Layer Deposition (ALD), are inclusions free and show homogeneous Mn distribution. Such films are paramagnetic with traces of anti-ferromagnetism, due to anti-ferromagnetic coupling of close TM-TM pairs [5,7]. Similar situation was observed by us for ZnCoO films grown by the ALD at similar conditions, as discussed elsewhere [10].

In this work we discuss optical (absorption spectra) and magneto-optical properties of ZnMnO samples grown by the ALD at low temperature (at 160 $^{o}$C) and with low Mn fractions (between 1 and 3%). Apparently contradicting results of previous magneto-optical investigations [11-13] will be explained here assuming Mn recharging in ZnMnO.

## 2. Experimental

### 2.1. Samples

The present optical and magneto-optical investigations were performed on ZnMnO thin films grown at LT (at 160 ºC) by the ALD. In the ALD we employed metalorganic zinc and manganese precursors: diethylzinc (DEZn) as zinc precursors and two types of Mn precursors: manganese-tris- 2,2,6,6-tetra methyl -3,5-heptanedione (denoted as $Mn(thd)_3$) and tris(2,4-pentanedionato) manganese(III) (denoted as $Mn(acac)_3$). As an oxygen precursor we used deionized water. Temperature of precursors were as follows: room temperature for DEZn, 150 – 160 ºC for $Mn(acac)_3$, and 150-180 $^0$C for $Mn(thd)_3$. Substrate ((001) Si (for magnetic investigations), (0001) sapphire (for optical absorption investigations) and glass (for electrical measurements)) temperature was 160 ºC. Pressure in a growth chamber, i.e., a pressure of $N_2$ transport and purging gas, was a few mbar. Ratio of Zn – to - Mn ALD cycles was either, 9 to 1 or 10 to 1, which we found to be optimal for a growth of fairly depth homogeneous ZnMnO films with low Mn fractions [5,7]. Some optical investigations were also performed for films grown with 10 to 5 Zn – to –Mn ALD cycles.

Thickness of our samples was between 0.6-0.8 µm. Most of the samples studied showed strong n-type conductivity, with free electron concentration (at RT) up to $10^{19}$ $cm^{-3}$ and mobility of a few $cm^2/Vs$. Samples with higher Mn content were more resistive. As-grown samples were investigated without any post-growth thermal treatment.

## 2.2 Experimental set ups

PL and magneto-PL investigations were performed at 2 K with samples mounted in a Spectromag 6000 split coil superconductive magnet system of Oxford Instruments. He-Cd laser PLASMA model HCCL-15UM was used for PL excitation. PL spectra were detected with a double monochromator LOMO MDR-23 equipped with Hamamatsu S7035 CCD camera and Hamamatsu photon counting system with R2531 photomultiplier (PMT) and FAST ComTec 7887 card. Further details on set ups used for PL and magneto-PL investigations can be found in the reference [11].

Investigations of absorption were performed at RT using either the Perkin Elmer Lambda 950 spectrophotometer or the Solar Spectrofluorimetr SM 2203 with two double monochromators, the Xe lamp as excitation source and the Hamamatsu PMT. The Hall effect measurements were performed in van der Pauw configuration in $B$=0.43T, using RH2035 system made by PhysTech GmbH Germany.

## 3. Experimental results
## 3.1 Magneto-optical investigations

The first puzzling results of PL and magneto-PL investigations of ZnMnO was that visible PL was not observed for most of the studied by us ZnMnO samples. We could detect visible PL only in samples containing fairly low Mn fraction (typically about 1 - 3 % and less). The 2K PL spectrum was dominated by neutral bound excitonic (DBE) emission with a weak free excitonic emission at high energy wing of the DBE peak. Deep defects related PL (in green-red spectral region) was either very weak or not observed, indicating good quality and stoichiometry of our films. For samples with low Mn fraction we measured PL and its circular polarization for magnetic field varied between 0 and 6 T. The results of relevant experiments for ALD-grown ZnMnO film with about 1% Mn fraction are shown in Fig. 1 (a, b), Fig. 2 and Fig. 3.

Figs. 1 and 2 show a strong PL polarization of the DBE emission observed for the ZnMnO film with 1 % Mn fraction. Similar PL response we observed for bulk ZnMnO sample, as reported previously [6]. Sigma + component increases in the intensity faster than sigma -, resulting in a strong PL circular polarization, with P = 0.35 defined as:

$$P = \frac{I_{\sigma+} - I_{\sigma-}}{I_{\sigma+} + I_{\sigma-}}$$

In common diluted magnetic semiconductor (DMS) strong PL polarization arises from large Zeeman splitting of excitonic transitions. Thus, large spectral shifts (by a few meV) and splitting were expected by us in the ZnMnO, due to strong magneto-optical effects commonly seen in DMS samples. Very surprisingly no spectral shift and resolved splitting were observed for excitonic lines in the present case, for magnetic field up to 6 T (see Fig. 3).

**3.2 Optical absorption investigations**

For ZnMnO samples we consistently observed an absorption band appearing below the onset of the band-to-band transition. This band, shown in Fig. 4, depends on Mn fraction in the sample, as we concluded from the optical absorption measurements performed for the ZnMnO layers with different Mn fractions [14]. The origin of this Mn-related absorption band is not clear. This band, first reported by Fukumura et al. [15] and tentatively related to charge transfer transitions between donor and acceptor ionization levels of Mn ions and the band continuum [15,16], was then related to smeared out the $Mn^{2+}$ intra-shell transitions [17,18].

In the case of ZnMnS (bulk sample, thin films and nanoparticles) we could resolve several intra-shell transitions [19]. Observation of these transitions in ZnMnS allowed us to verify critically the accuracy of the cluster model calculations presented in the reference [18]. Energy position of $^6A_1$ to $^4T_1$, $^4T_2$, $^4E$ and $^4A_1$ transitions were calculated for GaMnAs, ZnMnTe, ZnMnSe, ZnMnS, and for ZnMnO. The relevant values for ZnO are as follows (in eV): 2.55, 2.85, 2.97 and 2.99.

Even though the accuracy of these calculations is limited, the model predicted relatively well energies for ZnMnS. We assumed thus that the $^6A_1$ to $^4T_1$ transition in ZnMnO should occur at about 2.45 eV and that the one to $^4T_2$ level at 2.75 eV. No traces of these absorption bands are seen in Fig. 4. Instead, a broad absorption band of unknown origin dominates the below band gap absorption, as also observed in [15-18]. We will speculate further on that this broad absorption is due to a Mn-related photo-ionization transition. We speculate, since the direct proof is still missing, as explained below.

Photo-ionization character of a given absorption band can be proved based on photo-ESR investigations, using the method reviewed by us in [20]. Photo-ESR worked in the case of GaMnN samples [21], but failed in the present work. The postulated 3+ charge state of Mn ions was too short lived in our n-type samples to achieve the required population of this state for the ESR detection. We also noticed that light sensitivity of resistance of our films was very weak, due to their high initial n-type conductivity. We plan in future to perform such experiments in more resistive samples.

## 4. Discussion

Strong magneto-optical effects are commonly seen in DMS samples. For example, pronounced MCD signals were reported for ZnMnO films by Ando et al. [13] and were explained by the presence of strong sp-d mixing. In such the case, excitonic emission should shift towards lower energy (by few meV) and a split should be observed for PL measured at different polarizations, as was observed for p-type ZnMnO layers [12] and was reported by us for CdMnTe layers with a similar Mn concentration [22]. Thus, in our magneto-optical investigations of ZnMnO samples we looked for similar effects. Surprisingly, no spectral shifts and well-resolved splitting were observed for magnetic field up to 6 T.

There is a possible explanation of this discrepancy. For example, excitonic splitting was reported for p-type sample obtained by oxidation (by thermal treatment at 600-650 $^{o}$C in $O_2$) of ZnMnTe layers grown by MBE. Such samples (due to processing at high temperature) are very likely less uniform, should have regions of Mn accumulation and large potential fluctuations. In this case PL may come from regions with large Mn fractions and be dominated by recombination of localized excitons. In turn, in samples with larger Mn fractions, as the one studied by us previously [11], PL is quenched in Mn rich regions and comes only from regions with a reduced Mn concentration. If so, magneto-optical effects should be weak.

Such explanation of the data is supported by the results of our cathodoluminescence (CL) investigations of ZnMnO samples containing various Mn fractions (typically below 5 %) [23].

We observed large in-plane and depth fluctuations of the CL intensity related to nonuniform distribution of Mn in samples grown with higher Mn fraction or at higher temperature.

Since an inhomogeneous Mn distribution may affect results of magneto-optical investigations, we performed PL and magneto-PL investigations of samples with low Mn fractions grown at low temperature with the ALD, i.e., on samples with highly homogeneous Mn distribution, as reported previously [5,7].

To explain the puzzling results we supported magneto-optical investigations by measurements of PL and optical absorption. Mn doping stimulates a visible PL of ZnS [19]. It is not the case in ZnMnO. Here we observed that PL is deactivated in samples with larger Mn fractions and commonly was not observed for samples with Mn fraction above 5%. If observed for such samples, it was coming from regions with a reduced Mn concentration, as observed in the CL investigations [23], as mentioned above. A possible hint how to account for the PL quenching in ZnMnO comes from investigations of GaMnN. Multivalence of Mn ions in GaN lattice was quite convincingly proved [21,24-27]. $Mn^{2+}$ to $Mn^{3+}$ recharging was observed upon illumination in GaMnN [21]. It was also claimed that multivalence of Mn in GaMnN can influence magnetic properties of this material [24]. $Mn^{4+}$ to $Mn^{3+}$ recharging was also evidenced there [25,26], but its origin remains unclear. It may as well be due to existence of foreign phases in GaMnN [27]. Multivalence of Mn in ZnMnO was also suggested in [28]. This is why we looked for evidences of such process.

Let us assume the recharging of Mn ions in ZnO. If we attribute the observed absorption, with the onset at about 2.0 - 2.1 eV, to the 2+ to 3+ Mn ionization in ZnO, the results obtained by us, and also those reported by others, can be consistently interpreted. If Mn is ionized upon illumination we can easily explain why we do not see the $Mn^{2+}$ related intra-shell transitions, since these transitions overlap with the photo-ionization band.

Assuming the photo-ionization character of the absorption shown in Fig. 4, we could evaluate the relevant ionization energies. For this we used the approach proposed by Kopylov and Pihktin [29], which takes into account a lattice relaxation following a change of impurity charge state. Details of the modeling (dashed line in Fig 4) are given elsewhere [14]. Using the above model we got the following parameters for Mn 2+ to 3+ ionization ($E_{opt} = 2.6 \pm 0.1$ eV and $E_{th} = 2.1 \pm 0.1$ eV) [14].

If the band shown in Fig. 4 is due to Mn 2+ to 3+ ionization, we can explain many of puzzling ZnMnO properties, not only why the $Mn^{2+}$ intra-shell transitions are missing both in the absorption and in the PL experiments. We can account for the observed PL deactivation by Mn ions in ZnO lattice, as we observed in our previous CL investigations [23] and as reported by other authors. Possible explanation comes from our previous investigations of PL deactivation in ZnS [30,31] and ZnSe [32] doped with Fe impurities. "Killer action" of Fe ions on ZnS and ZnSe visible PL was the direct consequence of a mix valence of iron (2+ and 3+) and was related to the three mechanisms (see [33] and references given in): 1) to efficient carrier recombination via a mid band gap Fe-related level (the so-called bypassing effect [31]); 2) to efficient Auger-type energy transfer from excitons and DAPs to Fe ions, and; 3) to formation of complex centers of Fe with common PL activators in wide band gap II-VI materials. Two of the former processes (bypassing and Auger effect) are most likely responsible for the PL deactivation in the present case, assuming Mn ionization.

Discussion given above was based on the assumption of the mixed valence of Mn ions. This proposition is likely supported by the results of photoemission study, where Mn 3p states show two contributions with different binding energies separated by about 4 eV [34].

The presence of the two Mn charge states in ZnMnO may also account for the observed strong PL polarization. Similar strong PL polarization we observed recently for Cr doped ZnSe and ZnTe [35]. There a very strong PL polarization we related to spin selective carrier retrapping via Cr ions. Likely a similar process may account for the present results.

The relevant question remains are ZnMnO films suitable for spintronics applications, i.e., can we achieve RT FM in this material (not due to inclusions). The Zener's p-d exchange model proposed in the reference [2] will not work. If Mn changes its charge state, it will not stay in 2+ charge state in p-type samples. However, the character of recharging process is not clear at present. The process can be due to 2+ - to - 3+ photoionization ($Mn^{2+} + h\nu \rightarrow Mn^{3+} + e_{conduction\ band}$), as postulated here, or due to formation of so-called charge transfer state, an analogue of impurity bound exciton [36], with hole strongly localized at $Mn^{2+}$ center and with a loosely bound electron. Formation of such states was indirectly documented in case of several rare earth ions and also postulated for some of transition metal ions (see [36] and references given there).

Consequences of such Mn-related mid gap states for magnetic properties of ZnMnO were recently described in [37,38]. In particular, T. Dietl [37] discusses possible reasons for discrepancy between results of photoemission, XAS and various PL and magneto-optical investigations. He relates small exciton splitting in ZnMnO, as compared to e.g. CdMnTe and as reported here by us, to the presence of mid gap hole trap level due to Mn ions. His model underlies a key role of a hole localization for explanation of otherwise contradictory results. This model predicts also correctly the observed by us an increase of ZnMnO band gap with increasing Mn fractions, which was not described by commonly used mean field approximation.

**Conclusions**

Concluding, optical and magneto-optical investigations of ZnMnO favor the model of mix valence of Mn ions in ZnMnO. We can then easily account for the lack of Mn intra-shell transitions and for PL quenching, reported consistently in many reports. Such possibility is however a bad news for the initially proposed model of RT FM of ZnMnO. It looks very unlikely that we can simultaneously realize a required high concentration of Mn ions in 2+ charge state and achieve nearly metallic p-type conductivity. Hole localization at Mn-related mid gap level may explain many of contradictory experimental results, as also explained theoretically in two recent works [37,38].

**Acknowledgments**

This work was partly supported by grant no 1 P03B 090 30 of MEiN, Poland, by FunDMS ERC Advanced Grant Research and by the European Union within European Regional Development Fund, through grant Innovative Economy (POIG.01.01.02-00-008/08). M. Godlewski thanks CNRS for the financial support during his stay in Versailles.

**Figure captions:**

**Figure 1 (a, b):** Results of 2K magneto-optical investigations of ZnMnO layer with low Mn fraction (about 1 %). Changes of circular polarization of the band edge PL are shown for magnetic field varied between 0 and 6 T.

**Figure 2 (a, b):** Results of 2K magneto-optical investigations of ZnMnO layer with low Mn fraction (about 1 %). Changes in intensity of circular polarization of the DBE PL are shown in (a) for magnetic field varied between 0 and 6 T. In (b) we show the resulting PL polarization.

**Figure 3:** Results of 2K magneto-optical investigations of ZnMnO layer with low Mn fraction (about 1 %). Band edge PL for two circular polarizations is shown for magnetic field set at 6 T.

**Figure 4:** Room temperature absorption band of ZnMnO layer grown at low temperature with ALD with about 3 % Mn fraction and with uniform Mn distribution. Dashed line represents fit to the experimental data open circles) with model taking into account lattice relaxation.

**Figure 1 (a, b)**

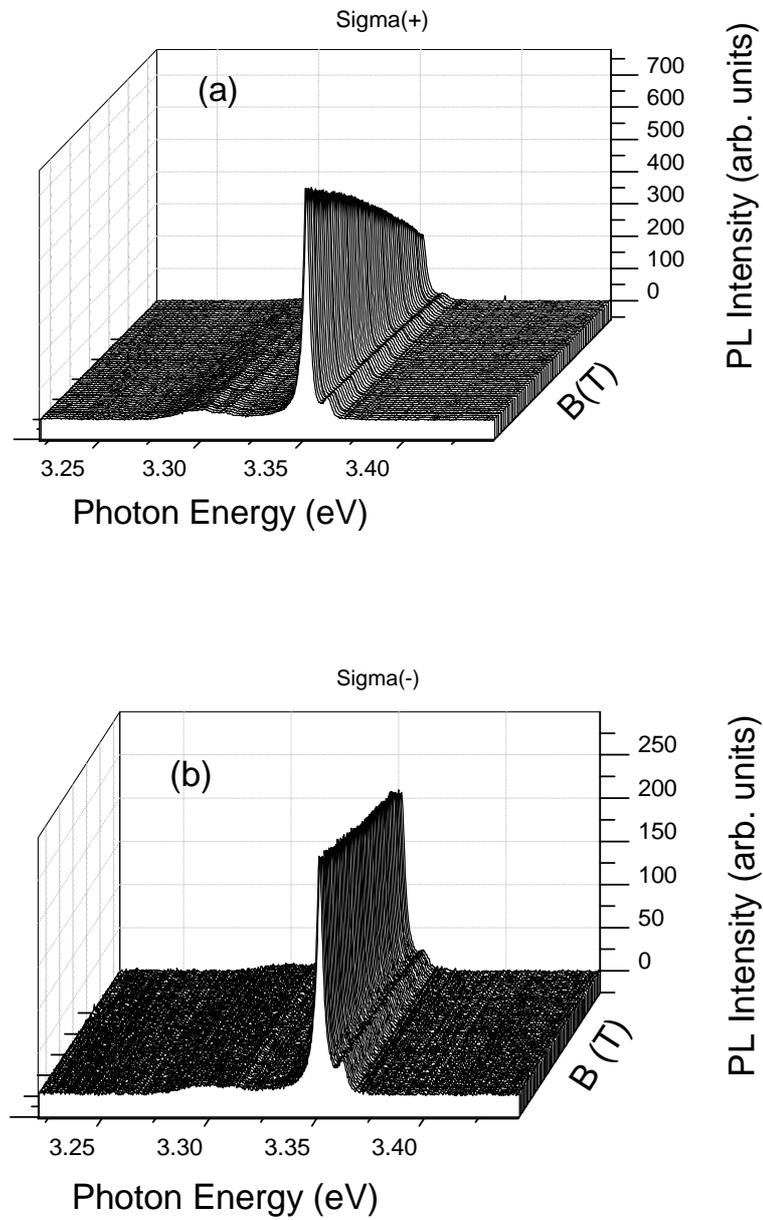

**Figure 1 (a, b):** Results of 2K magneto-optical investigations of ZnMnO layer with low Mn fraction (about 1 %). Changes of circular polarization of the band edge PL are shown for magnetic field varied between 0 and 6 T.

**Figure 2 (a, b)**

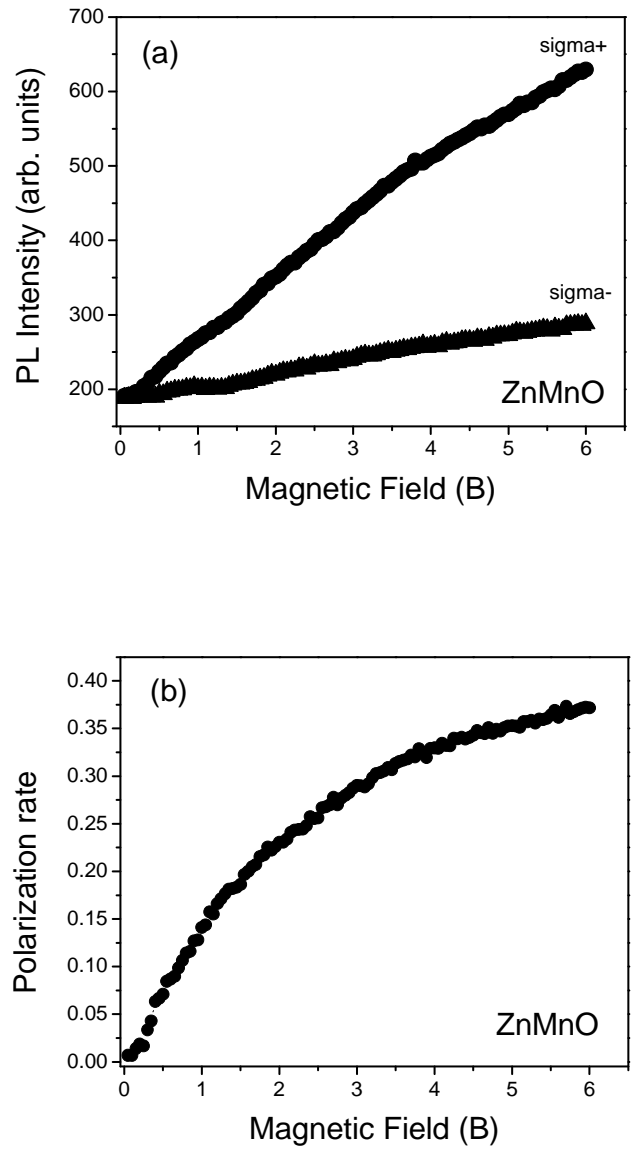

**Figure 2:** Results of 2K magneto-optical investigations of ZnMnO layer with low Mn fraction (about 1 %). Changes in intensity of circular polarization of the DBE PL are shown in (a) for magnetic field varied between 0 and 6 T. In (b) we show the resulting PL polarization.

**Figure 3:**

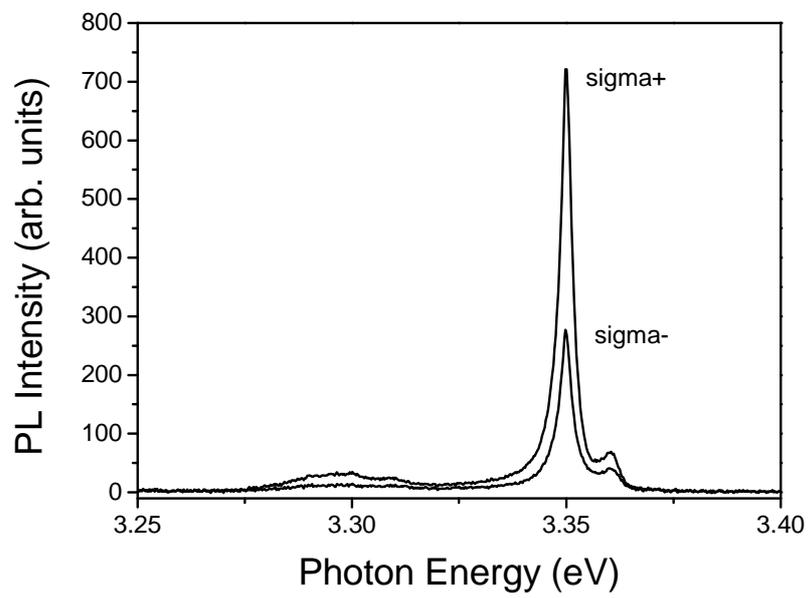

**Figure 3**: Results of 2K magneto-optical investigations of ZnMnO layer with low Mn fraction (about 1 %). Band edge PL for two circular polarizations is shown for magnetic field set at 6 T.

**Figure 4:**

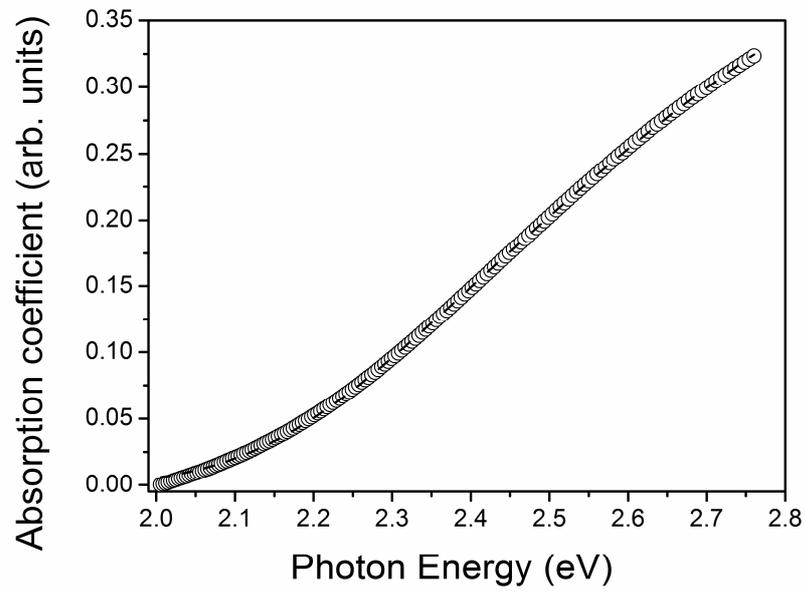

**Figure 4:** Room temperature absorption band of ZnMnO layer grown at low temperature with ALD with about 3 % Mn fraction and with uniform Mn distribution. Dashed line represents fit to the experimental data open circles) with model taking into account lattice relaxation.